\documentclass[eqsecnum,twocolumn,aps,epsf]{revtex4}
\usepackage{graphicx}
\usepackage{color}
\newcommand{\Tsub}[2]{\ensuremath{#1_{\mathrm{#2}}}}

\newcommand{\Tk}[1]{\Tsub{\kappa}{#1}}

\begin{document}
\draft

\title{Thermal Conductivity of the Quasi One-Dimensional Spin System Sr$_2$V$_3$O$_9$}

\author{M. Uesaka}
\address{Department of Applied Physics, Tohoku University, 6-6-05 Aoba, Aramaki, Aoba-ku, Sendai 980-8579, Japan}
\thanks{Corresponding author: uesaka@teion.apph.tohoku.ac.jp}
\author{T. Kawamata}
\address{Nishina Center for Accelerator-Based Science, RIKEN, 2-1 Hirosawa, Wako 351-0198, Japan}
\author{N. Kaneko, M. Sato}
\address{Department of Applied Physics, Tohoku University, 6-6-05 Aoba, Aramaki, Aoba-ku, Sendai 980-8579, Japan}
\author{K. Kudo, N. Kobayashi}
\address{Institute for Materials Research, Tohoku University, 2-1-1 Katahira, Aoba-ku, Sendai 980-8577, Japan}
\author{Y. Koike}
\address{Department of Applied Physics, Tohoku University, 6-6-05 Aoba, Aramaki, Aoba-ku, Sendai 980-8579, Japan}

\date{\today}

\begin{abstract}

We have measured the thermal conductivity along the [101] direction, \Tk{[101]}, along the [10$\bar{1}$] direction, \Tk{[10\bar{1}]}, and along the $b$-axis, $\kappa_\mathrm{b}$, of the quasi one-dimensional $S$=1/2 spin system Sr$_2$V$_3$O$_9$ in magnetic fields up to 14 T, in order to find the thermal conductivity due to spinons and to clarify whether the spin-chains run along the [101] or $[10\bar{1}]$ direction. 
It has been found that both \Tk{[101]}, \Tk{[10\bar{1}]} and $\kappa_\mathrm{b}$ show one peak around 10 K in zero field and that the magnitude of \Tk{[10\bar{1}]} is larger than those of \Tk{[101]} and $\kappa_\mathrm{b}$. 
By the application of magnetic field along the heat current, the peak of $\kappa_{[10\bar{1}]}$ is markedly suppressed, while the peaks of $\kappa_{[101]}$ and $\kappa_\mathrm{b}$ little change. 
These results indicate that there is a large contribution of spinons to $\kappa_{[10\bar{1}]}$ and suggest that the spin-chains run along the [10$\bar{1}$] direction.

\end{abstract}
\vspace*{2em}
\maketitle
\newpage

\section{Introduction}\label{intro}
Recently, the thermal conductivity in low-dimensional quantum spin systems has attracted interest, 
because a large contribution of magnetic excitations to the thermal conductivity has been observed in some compounds regarded as low-dimensional quantum spin systems. 
In one-dimensional (1D) antiferromagnetic (AF) Heisenberg spin systems with the spin quantum number $S$ = 1/2, especially, it has theoretically been proposed that the thermal conduction due to magnetic excitations, namely, spinons is ballistic \cite{Castella:PRL74:1995:972,Zotos:PRB55:1997:11029,Klumper:JPA35:2002:2173}. 
In fact, a large contribution of spinons to the thermal conductivity has been observed in the $S$=1/2 1D AF Heisenberg spin system Sr$_2$CuO$_3$ \cite{Sologubenko:PRB62:2000:R6108}. 
Moreover, the ballistic nature of the thermal conduction due to spinons has experimentally been confirmed in Sr$_2$CuO$_3$ \cite{Kawamata:JPSJ77:2008:034607}.

\begin{figure}[tbp]
\begin{center}
\includegraphics[width=15pc]{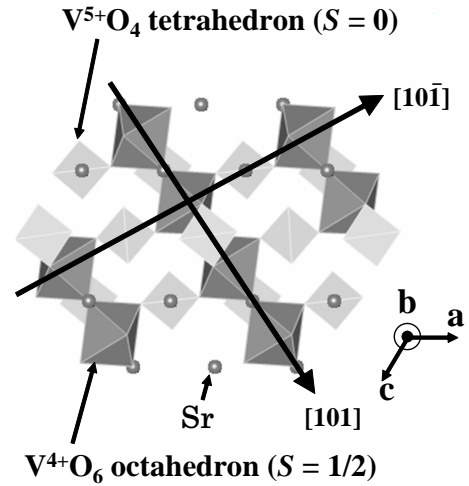}
\caption{Crystal structure of Sr$_2$V$_3$O$_9$. V$^{5+}$ ions and V$^{4+}$ ions are located in VO$_4$ tetrahedra and VO$_6$ octahedra, respectively. VO$_6$ octahedra are connected with each other along the [101] direction by sharing an oxygen at the corner and are also connected via a VO$_4$ tetrahedron along the [10$\bar{1}$] direction.}
\label{kouzou}
\end{center}
\end{figure}

The compound Sr$_2$V$_3$O$_9$ contains three kinds of vanadium ions in the unit cell. Two of them are nonmagnetic V$^{5+}$ ions located in VO$_4$ tetrahedra, and the rest is V$^{4+}$ ions with $S = 1/2$ located in VO$_6$ octahedra. The VO$_6$ octahedra are connected with each other by sharing an oxygen at the corner along the [101] direction, as shown Fig. \ref{kouzou}. 
The $ac$-plane with the magnetic network is weakly stacked along the $b$-axis. 
The magnetic properties are understood in terms of the $S$=1/2 1D AF chain model with the exchange interaction between the nearest spins, $J = 82$ K, estimated from the susceptibility measurements \cite{Kaul:PRB67:2003:174417}.
 However, it has been suggested from the ESR measurements 
 \cite{Ivanshin:PRB68:2003:064404}
 and a theory by Koo and Whangbo 
 \cite{Koo:SSS9:2007:824}
 that the spin-chain direction is not the [101] direction but the [10$\bar{1}$] direction, along which VO$_6$ octahedra are connected via a VO$_4$ tetrahedron as shown Fig. \ref{kouzou}. 

Therefore, in order to find the contribution of spinons to the thermal conductivity and also to clarify whether the spin-chain direction is the [101] or [10$\bar{1}$] direction, we have measured the thermal conductivity of Sr$_2$V$_3$O$_9$ along the [101], [10$\bar{1}$] directions and the $b$-axis. 
The precise measurement of thermal conductivity needs a large-size single crystal. 
Therefore, we have attempted to grow large-size single crystals of Sr$_2$V$_3$O$_9$ by the floating-zone (FZ) method.

\section{Experimental}
 First, polycrystalline powder of Sr$_2$V$_2$O$_7$ was prepared by the solid-state reaction method, in order to prepared a feed rod for the FZ growth.
 The prescribed amount of SrCO$_3$ and V$_2$O$_5$ powder was mixed in the molar ratio of SrCO$_3$ : V$_2$O$_5$ = 2 : 1 and prefired in air at 700\ensuremath{\mathrm{\char'27 \kern-.2em \hbox{C}}} for 72 h.
 After pulverization, the prefired powder of Sr$_2$V$_2$O$_7$ was mixed with VO$_2$ powder in the molar ratio of Sr$_2$V$_2$O$_7$ : VO$_2$ = 1 : 1 and isostatically cold-pressed at 600 bar into a rod of 7 mm in diameter and $\sim$100 mm in length. Then, the rod was sintered at 540\ensuremath{\mathrm{\char'27 \kern-.2em \hbox{C}}} in Ar for 24 h.
 As a result, a sintered feed rod was prepared. The FZ growth was carried out by the twice-scanning technique in an infrared heating furnace equipped with a double ellipsoidal mirror (NEC Machinery Corp, Model SC-K15HD-H).
 A high-density premelted feed rod was prepared through the first scan using the sintered feed rod.
 In the first scan, the molten zone was scanned at a speed of $\sim$20 mm/h under flowing Ar of 1.5 atm.
 Next, the second scan, namely, a usual growing procedure was carried out using the premelted feed rod at the growth rate of 1.0 mm/h in the same atmosphere as in the first scan.
 Thermal-conductivity measurements were carried out by the conventional steady-state method.

\section{Results and Discussion}
\begin{figure}[tbp]
\begin{center}
\includegraphics[width=13pc]{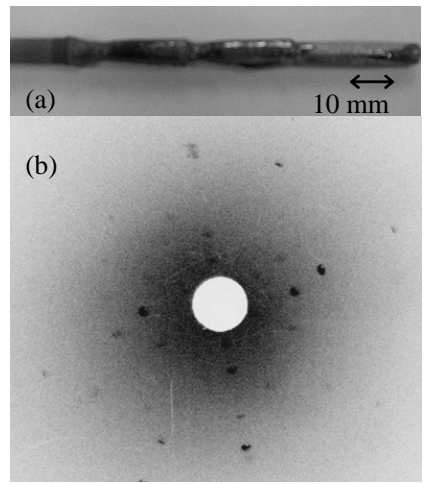}
\caption{(a) Picture of an as-grown single-crystal rod of Sr$_2$V$_3$O$_9$.	(b) X-ray back-Laue photography of an as-grown single-crystal in the x-ray parallel to the $b$-axis.}
\label{crystal}
\end{center}
\end{figure}

We have succeeded in growing a single-crystal rod, owing to the stable upkeep of the molten zone during the FZ growth. 
Figure 2(a) shows an as-grown single-crystal rod with $\sim$ 6 mm in diameter and $\sim$ 100 mm in length. 
The grown crystals were characterized by the x-ray back-Laue photography, as shown in Fig. 2(b).
Although the grown crystals were composed of several domains, the diffraction spots were very sharp.
The dimensions of the single-domain region were typically $\sim$ 6 mm in diameter and $\sim$ 25 mm in length.
The single crystals were also confirmed by the powder x-ray diffraction to be of the single phase without any impurity phases.
Accordingly, it is concluded that we have succeeded in the growth of high-quality single-crystals. 
The high quality was supported by the magnetic-susceptibility result that no Curie term due to impurities and/or lattice defects was observed at very low temperatures. 

\begin{figure}[tbp]
\begin{center}
\includegraphics[width=18pc]{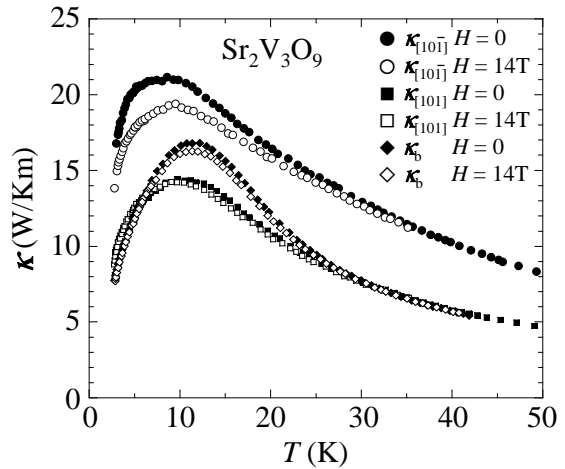}
\caption{Temperature dependence of the thermal conductivity of Sr$_2$V$_3$O$_9$ along the [101] direction, \Tk{[101]}, along the [10$\bar{1}$] direction, \Tk{[10\bar{1}]}, and along the $b$-axis, $\kappa_\mathrm{b}$, in zero field and a magnetic field of 14 T parallel to the respective heat current.}
\label{fig:netu2}
\end{center}
\end{figure}

\begin{figure}[tbp]
\begin{center}
\includegraphics[width=18pc]{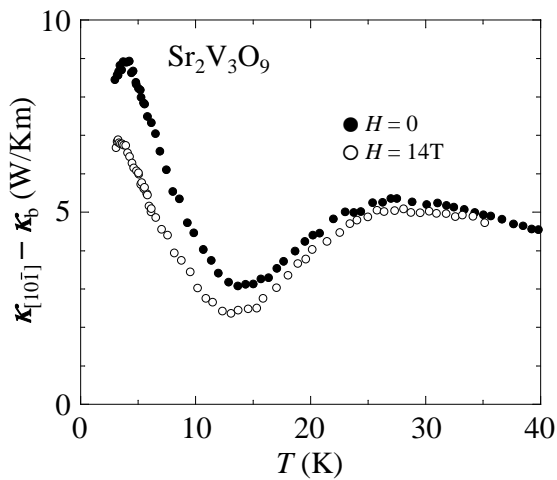}
\caption{Temperature dependence of the difference between $\kappa_{[10\bar{1}]}$  and $\kappa_\mathrm{b}$ in zero field and 14 T. }
\label{fig:netu2}
\end{center}
\end{figure}

Figure 3 shows the temperature dependence of the thermal conductivity 
along the [101] direction, \Tk{[101]}, along the [10$\bar{1}$] direction, \Tk{[10\bar{1}]}, and along the $b$-axis, $\kappa_\mathrm{b}$, in zero field and a magnetic field of 14 T parallel to the respective heat current.
In zero field, both \Tk{[101]}, \Tk{[10\bar{1}]} and $\kappa_\mathrm{b}$ show a peak around 10 K. 
The magnitude of \Tk{[10\bar{1}]} at the peak is larger than those of \Tk{[101]} and $\kappa_\mathrm{b}$. 
Since Sr$_2$V$_3$O$_9$ is an insulator, the thermal conductivity is described as the sum of the thermal conductivity due to phonons, \Tk{phonon}, and due to spinons, \Tk{spinon}. 
It is known that the anisotropy of \Tk{phonon} is usually not so large and that the contribution of \Tk{spinon} markedly appears along the direction where the magnetic interaction is strong. 
Therefore, the large anisotropy of the thermal conductivity is guessed to be due to a large contribution of \Tk{spinon} to \Tk{[10\bar{1}]}. 
By the application of magnetic field parallel to the heat current, the peak of \Tk{[10\bar{1}]} around 10 K is suppressed with increasing field, 
while there is little change in \Tk{[101]} and $\kappa_\mathrm{b}$, as shown in Fig. 3. 
This result also supports the guess that there is a large contribution of \Tk{spinon} to \Tk{[10\bar{1}]}, 
because \Tk{spinon} is expected to be affected by the application of a magnetic field comparable with $J/(g \mu _{\mathrm{B}})$ ($g$: the $g$-factor, $\mu _{\mathrm{B}}$: the Bohr magneton). 
Furthermore, the little change in \Tk{[101]} and $\kappa_\mathrm{b}$ by the application of magnetic field indicates that the contribution of \Tk{spinon} is very small along the [101] direction and $b$-axis. 
Accordingly, it is concluded that the spin-chain direction is the [10$\bar{1}$] direction, as suggested from the ESR measurements 
\cite{Ivanshin:PRB68:2003:064404}
 and the theory by Koo and Whangbo 
 \cite{Koo:SSS9:2007:824}.
 
Here, we estimate $\Tk{spinon}$ in $\Tk{[10\bar{1}]}$, where both $\Tk{spinon}$ and $\Tk{phonon}$ are included.
In the temperature dependence of $\Tk{[10\bar{1}]}$, only one peak is observed around 10 K, indicating that both peaks of $\Tk{spinon}$ and $\Tk{phonon}$ are overlapped. Therefore, it is very hard to estimate $\Tk{spinon}$ and $\Tk{phonon}$ separately.
As for $\Tk{[101]}$, a small contribution of $\Tk{spinon}$ to $\Tk{[101]}$ is guessed to exist, because the [101] direction is not exactly perpendicular to the [10$\bar{1}$] direction but 84.48$\mathrm{\char'27 \kern-.2em}$ tilted from the [10$\bar{1}$] direction. As for $\kappa_\mathrm{b}$, it is due to only $\Tk{phonon}$. 
Therefore, neglecting the anisotropy of $\Tk{phonon}$, $\Tk{spinon}$ along the [10$\bar{1}$] direction is very roughly estimated to be $\Tk{[10\bar{1}]} - \kappa_\mathrm{b}$, as shown in Fig. 4.
However, this is not simply accepted as the temperature dependence of $\Tk{spinon}$ along the [10$\bar{1}$] direction, because unusual two peaks appear around 4 K and 28 K.
What is remarkable at least is that the peak around 4 K is strongly suppressed by the application of magnetic field while the other peak around 28 K little changes. 
Therefore, it is expected that the peak around 4 K is attributed to the contribution of \Tk{spinon}. 
On the other hand, it is likely that the peak around 28 K appears because of the difference of $\Tk{phonon}$ between the[10$\bar{1}$] direction and the $b$-axis.
Accordingly, at least these results indicate that the temperature dependence of \Tk{spinon} exhibits a peak around 4 K in \Tk{[10\bar{1}]}. 
In order to estimate the value of \Tk{spinon} in Sr$_2$V$_3$O$_9$ exactly, the estimate of the anisotropy of \Tk{phonon} between the [10$\bar{1}$] direction and $b$-axis is necessary.
\section{Summary}
Large-size single-crystals of Sr$_2$V$_3$O$_9$ have successfully been grown by the FZ method and the thermal conductivity have been measured in magnetic fields up to 14 T.
The magnitude of \Tk{[10\bar{1}]} in zero field is larger than those of \Tk{[101]} and $\kappa _\mathrm{b}$.
By the application of magnetic field, only \Tk{[10\bar{1}]} is suppressed. 
These anisotropic behaviors suggest that the spin-chains run along the [10$\bar{1}$] direction. 
Moreover, it is concluded from the field effect of $\Tk{[10\bar{1}]} - \kappa_\mathrm{b}$ related to the behavior of \Tk{spinon} that the temperature dependence of \Tk{spinon} exhibits a peak around 4 K. 
\section*{Acknowledgments}
The thermal conductivity measurements were performed at the High Field Laboratory for Superconducting Materials, Institute for Materials Research, Tohoku University. 
This work was partly supported by a Grant-in-Aid for Scientific Research from the Ministry of Education, Culture, Sports, Science and Technology, Japan.

\end{document}